\documentclass[conference]{IEEEtran}
\IEEEoverridecommandlockouts
\usepackage{cite}
\usepackage{amsmath,amssymb,amsfonts}
\usepackage{algorithmic}
\usepackage{graphicx}
\usepackage{textcomp}
\usepackage{xcolor}

\def\BibTeX{{\rm B\kern-.05em{\sc i\kern-.025em b}\kern-.08em
    T\kern-.1667em\lower.7ex\hbox{E}\kern-.125emX}}
    
\begin{document}

\title{Hybrid Anomaly Detection for Bullion Coin Authentication Leveraging Acoustic Signature Analysis}

\author{
	\IEEEauthorblockN{1\textsuperscript{st} Krzysztof Siwek}
	\IEEEauthorblockA{\textit{Warsaw University of Technology} 
		\\ Warsaw, Poland\\
		 ORCID: 0000-0003-2642-2319 \\
		 krzysztof.siwek@pw.edu.pl}
    \and
	\IEEEauthorblockN{2\textsuperscript{nd} Tran Hoai Linh}
	\IEEEauthorblockA{\textit{Hanoi University of Science and Technology}\\ 
    Hanoi, Vietnam \\
		ORCID: 0000-0001-9757-8041 \\
		linh.tranhoai@hust.edu.vn}
	\and
	\IEEEauthorblockN{3\textsuperscript{rd} Tomasz Gryczka}
	\IEEEauthorblockA{\textit{Student of} \\
		\textit{Warsaw University of Technology}\\
		 Warsaw, Poland }
	\and
	\IEEEauthorblockN{4\textsuperscript{th} Maciej Stodolski}
	\IEEEauthorblockA{\textit{Warsaw University of Technology} \\
		Warsaw, Poland \\
		ORCID: 0009-0003-0644-8105 \\
		maciej.stodolski@pw.edu.pl}
}

\maketitle

\begin{abstract}
The verification of bullion coin authenticity is essential for maintaining integrity within the precious metals market; however, the increasing sophistication of counterfeits has rendered traditional inspection methods insufficient. This paper proposes a non-destructive verification framework based on acoustic frequency analysis and deep neural networks. The methodology leverages the unique acoustic fingerprint of a coin, a physical signature determined by its material composition, mass, and geometry, captured through mechanical excitation. We implement a synergistic dual-model architecture consisting of an autoencoder that reconstructs the spectrum for anomaly detection and a deep learning classifier for coin type identification. To address the challenges of environmental noise and limited dataset diversity, a dynamically calculated anomaly threshold and data augmentation techniques were employed. Experimental results demonstrate that the integrated system achieves high precision in distinguishing authentic specimens from high-quality counterfeits, maintaining stability across varying recording conditions and devices. Beyond bullion authentication, the study highlights the scalability of the proposed non-destructive testing method for assessing the safety of critical components in the automotive and aerospace industries.
\end{abstract}


\section{Introduction}
Since the dawn of the first civilizations, the exchange of goods has relied on materials with relatively stable value, such as grain, salt, or ore. However, these early media suffered from significant drawbacks, they were often perishable, cumbersome to transport, and difficult to standardize. Over time, precious metals, specifically gold and silver, emerged as the superior choice due to their durability, divisibility, and ease of transport. 

The transition to precious metals as a primary currency took various forms, but the true turning point in monetary history was the introduction of coins \cite{BM96Walrasian, Dav94History}. These standardized units of weight and purity featured increasingly complex imagery, symbols, and graphic elements on their surfaces. This standardization reduced the need for constant, exhaustive testing of the payment medium during every transaction, allowing coins to become a universal measure of value and a reliable method for storing wealth. 

\subsection{The persistent challenge of counterfeiting}

Unfortunately, as coins became the dominant form of payment, counterfeiting followed almost immediately. By the Middle Ages, as trade expanded, forgery became a systemic problem. Counterfeiters employed a variety of techniques to mimic the appearance of genuine currency while minimizing actual precious metal content. Common methods included mixing precious metals with cheaper alternatives such as copper (debasement), casting coins from inexpensive lead or copper alloys, and covering them with a thin layer of gold or silver (plating). 

Despite severe legal penalties, often including the death penalty, counterfeit money continued to circulate. This ongoing threat to the monetary system forced the development of early verification methods, such as weighing, measuring dimensions, and visual inspections.

\begin{figure}[ht]
\centering
\begin{minipage}[b]{0.24\textwidth}
    \centering
    \includegraphics[width=\linewidth]{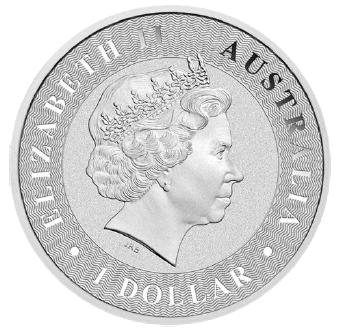}
    reverse
\end{minipage}\hfill
\begin{minipage}[b]{0.23\textwidth}
    \centering
    \includegraphics[width=\linewidth]{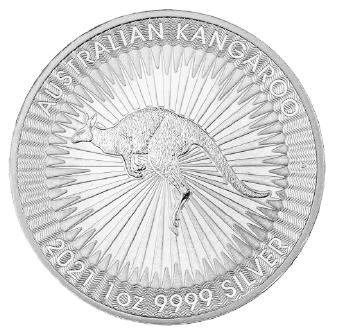}
    obverse
\end{minipage}
\caption{One-ounce Australian Kangaroo Silver Coin}
\end{figure}

More advanced techniques included the displacement method based on Archimedes' Principle, which allowed assessment of material density but required specialized tools and knowledge to execute \cite{Hug06Archimedes}.

\subsection{Sound analysis for verification}

One of the most enduring and intuitive verification techniques is the "ping test": listening to the sound a coin makes when dropped or struck against a hard surface. A genuine coin produces a clear, bell-like ring, whereas counterfeits made from inferior alloys often sound dull or muffled. Historically, this method relied entirely on the experience of the merchant, whose ear had to be finely tuned to the specific resonance of authentic coinage \cite{Suz08Acoustic}. However, as a standalone technique, it lacked the precision necessary to detect high-quality, modern counterfeit (fake) coins.

Modern technology has transformed this intuitive practice into a rigorous scientific discipline. Advances in microphone sensitivity and signal processing algorithms now enable precise recording and analysis of a coin's acoustic signature. Every coin has a unique sound-frequency profile defined by its composition, mass, and shape \cite{AS76Plates, DKG16Eigen}. 

\subsection{Neural networks for bullion authentication}

Bullion coins (also known as investment coins) are ideal candidates for this type of acoustic analysis. Typically minted from high-purity gold, silver, or platinum, their value is derived from their metal content rather than their face value or numismatic rarity. Because bullion markets are often perceived as stable stores of value during periods of economic uncertainty, preserving market confidence is essential. In addition, modern bullion coins are manufactured with exceptional precision, resulting in negligible differences in mass or geometry between authenticated and counterfeit specimens \cite{CK21GoldBullion}.

The spectrogram shown in the figure was created by striking a coin with a hard object. The coin was held on both sides at a point on the axis of rotation passing through the center. Analysis of the spectrogram reveals several distinct peaks at resonant frequencies, located primarily in the lower part of the spectrum, i.e., below 10 kHz. These peaks are located around 3500 Hz, 3750 Hz, 8700 Hz, and 15500 Hz. The resonant peaks appear as horizontal, brighter lines and are a symptom of sustained oscillations at these frequencies after the initial strike \cite{Boa16Signal}. 

The intensity of the resonant modes decreases over time, as seen in the gradual, slow disappearance of the bright horizontal lines towards the right of the spectrogram. Immediately after the strike, the highest signal energy is observed (approximately 0.1s), where the entire frequency range is excited by the strike. After the impact pulse fades, only resonant modes dominate the spectrogram. This behavior is typical of the natural vibrations of coins, where the resonance frequencies correspond to the natural vibration frequencies, which in turn result from the coin's material properties and shape.

\begin{figure}
    \centering
    \includegraphics[width=1\linewidth]{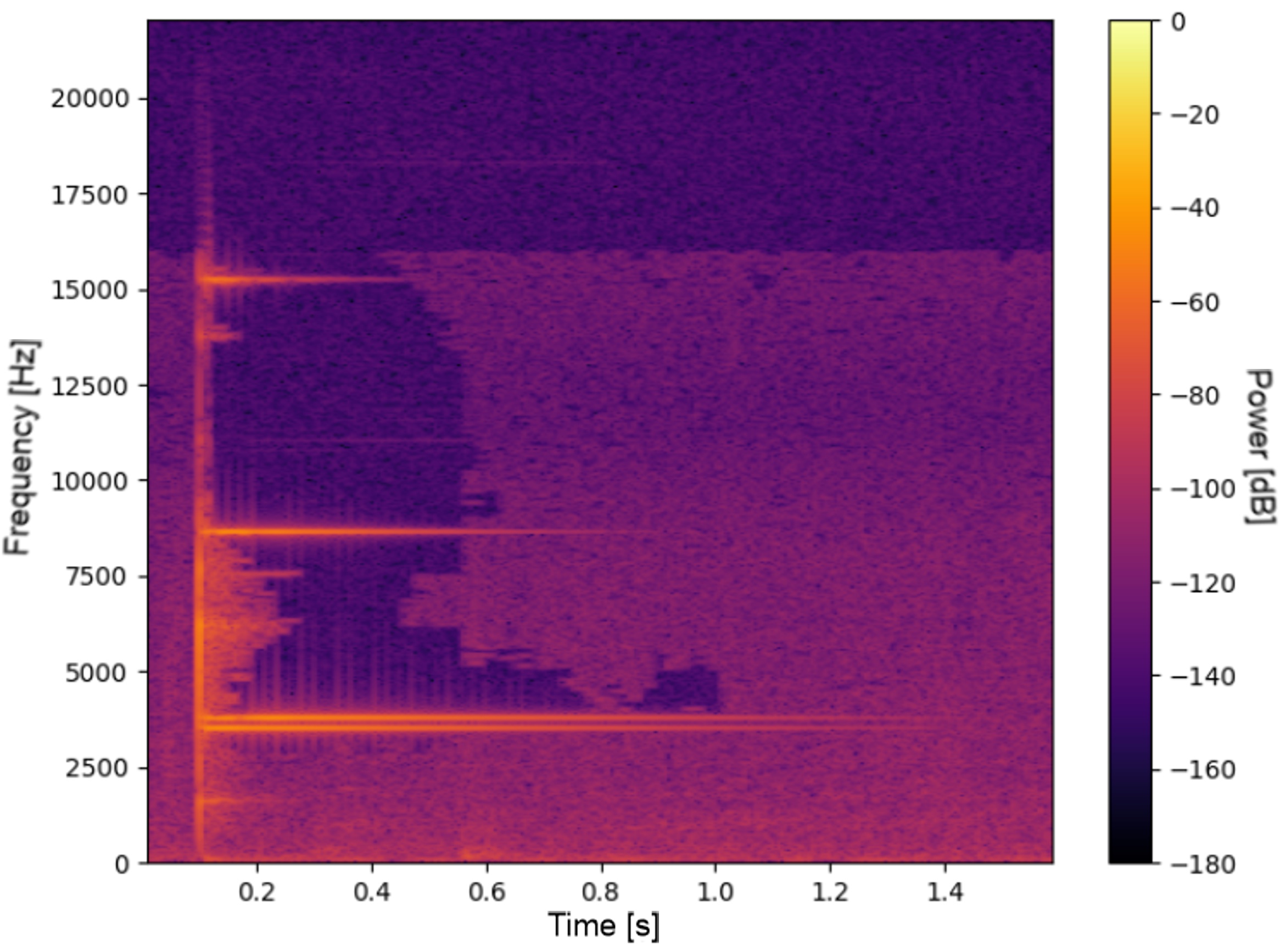}
    \caption{Mechanical impact spectrogram}
    \label{img_spectgrm}
\end{figure}

By leveraging Artificial Neural Networks (ANNs) to process these acoustic signals, we can move beyond human subjectivity. ANNs are uniquely capable of identifying the complex, nonlinear patterns within a coin's frequency spectrum that indicate its true material integrity. This research aims to develop a fast, simple, and effective method for bullion verification, providing investors with a robust tool to navigate the precious metals market with certainty.

\section{Metodology}

At the foundation of this methodology is treating acoustic response as a principal means of assessing a coin's physical characteristics. The recorded sound is considered an intrinsic acoustic signature, distinctive to each coin type. In this study, the acoustic signal is produced by mechanically striking the coin with a rigid object. This external excitation induces the coin's natural vibrational modes, which are directly governed by its metallurgical composition and geometric accuracy. By capturing and analyzing these vibrations, the system obtains high‑resolution data that forms the basis for reliable coin verification.

\begin{table}[ht]
\centering
\caption{Resonance peaks in the acoustic signal of the 1 oz Australian Kangaroo silver coin}
\begin{tabular}{c c c}
\hline
\textbf{Peak number} & \textbf{Frequency [Hz]} & \textbf{Relative Amplitude [dB]} \\
\hline
1 & 3770.00  & 0.00   \\
2 & 3505.62  & -0.36  \\
3 & 8648.75  & -6.78  \\
4 & 15258.12 & -22.18 \\
\hline
\end{tabular}
\label{tab1}
\end{table}

Table \ref{tab1} shows the observed resonance frequencies in the acoustic signal, which were extracted using the Fourier transform (FFT) from the signal shown in the previous spectrogram. By detecting local maxima in the power spectrum, accounting for minimum peak separation in the frequency domain, and applying a prominence threshold, it was possible to extract only the dominant resonance peaks from the signal \cite{OS97Discrete}.

The analysis revealed (as seen in the spectrogram) the presence of four resonance frequencies, with the dominant peak around 3770.00 Hz (0.00 dB), suggesting a fundamental natural vibration mode of the coin, which is typical of its material and geometric properties. Subsequent peaks with lower amplitudes correspond to higher harmonics or higher resonance modes, whose energy decreases with increasing frequency, consistent with the theory of vibration damping in metallic materials \cite{RSK17Damping}.

\begin{table}[ht]
\centering
\caption{extracted resonance frequencies of three different specimens of the 1 oz Australian Kangaroo silver coin}
\begin{tabular}{c c c c c}
\hline
\textbf{Modes} &
\textbf{Freq (Hz)} &
\(\sigma\)\textbf{Freq (Hz)} &
\textbf{Amp. (dB)} &
\(\sigma\)\textbf{Amp. (dB)} \\
\hline
1 & 3479.58 & 55.07 & -15.34 & 17.45 \\
2 & 3732.29 & 47.28 & -15.50 & 15.69 \\
3 & 8644.17 & 5.62  & -4.86  & 3.58  \\
4 & 15238.33 & 7.33 & -10.00 & 2.75  \\
\hline
\end{tabular}
\label{tab2}
\end{table}

Table \ref{tab2} presents the extracted resonance frequencies of three different specimens of the 1 oz Australian Kangaroo silver coin \cite{MK21Awers}. For each mode, the mean resonance frequency and relative amplitudes are reported, along with their standard deviations ($\sigma$), allowing assessment of inter-specimen variability.

The analysis demonstrated high uniformity among the study's specimen coins. In particular, modes in the upper frequency range (e.g., mode 3 at 8644 Hz, with a frequency standard deviation of 5.62 Hz and an amplitude of 3.58 dB) showed small deviations from the mean. The lower modes (1 and 2) exhibit slightly higher variability (frequency standard deviation is approximately 47–55 Hz and amplitude standard deviation between 15–18 dB), which can be attributed to differences in the striking method, vibration damping, recording device properties, as well as minimal differences in the manufacturing process and wear of a particular coin.

\begin{figure}
    \centering
    \includegraphics[width=1\linewidth]{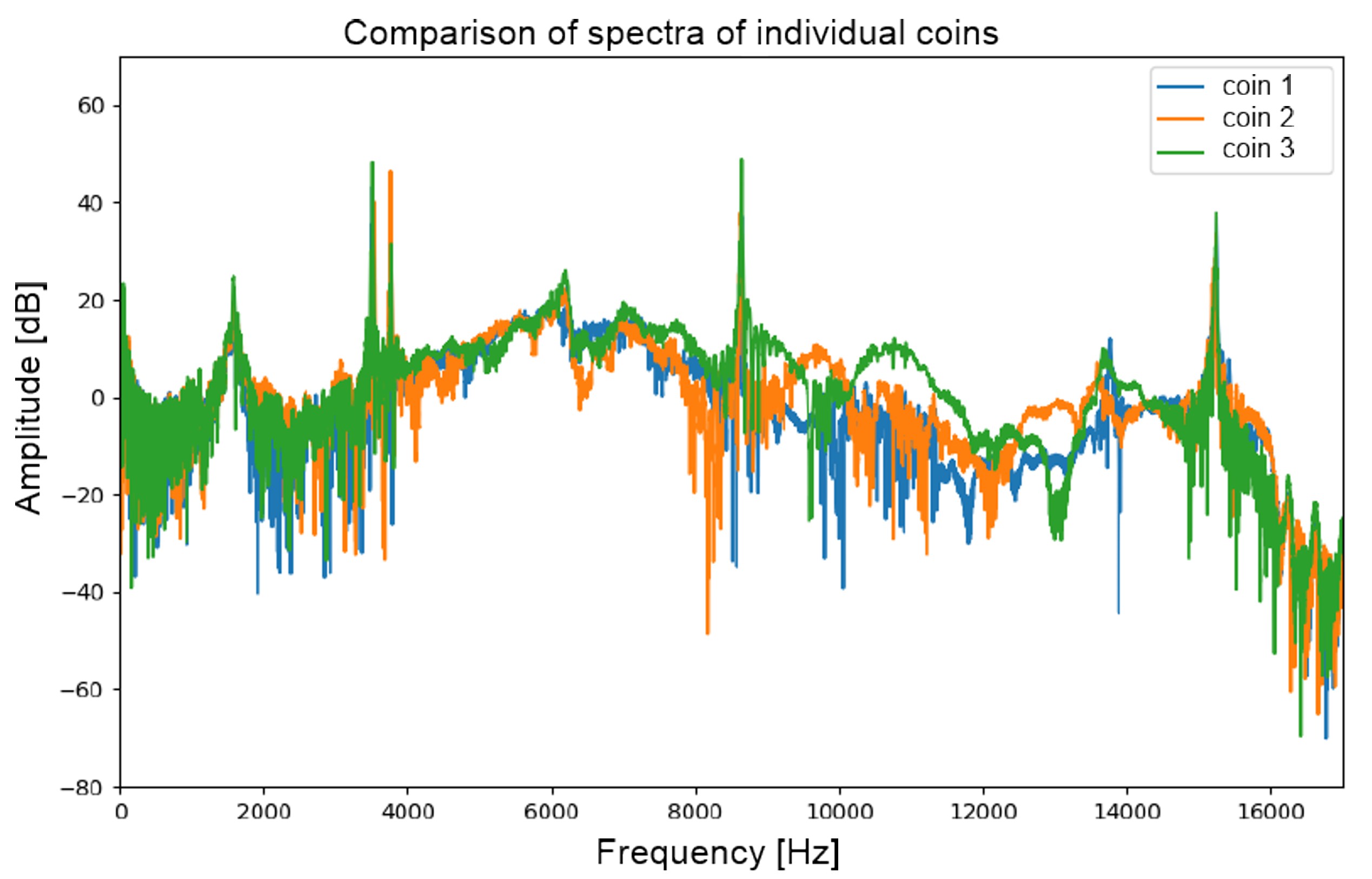}
    \caption{Spectrum of frequency spectra for three different specimens of the same bullion coin}
    \label{img_spectra_of3}
\end{figure}

Figure \ref{img_spectra_of3} illustrates the distribution of resonance peak positions across the full spectral range. The observed low standard deviation values support the assumption of specimen homogeneity, thereby providing a foundation for subsequent investigations into counterfeit detection based on acoustic resonance verification. Nevertheless, the spectral differences identified are insufficient to enable straightforward discrimination between individual specimens.

\subsection{Counterfeit spectra}

\begin{figure}
    \centering
    \includegraphics[width=1\linewidth]{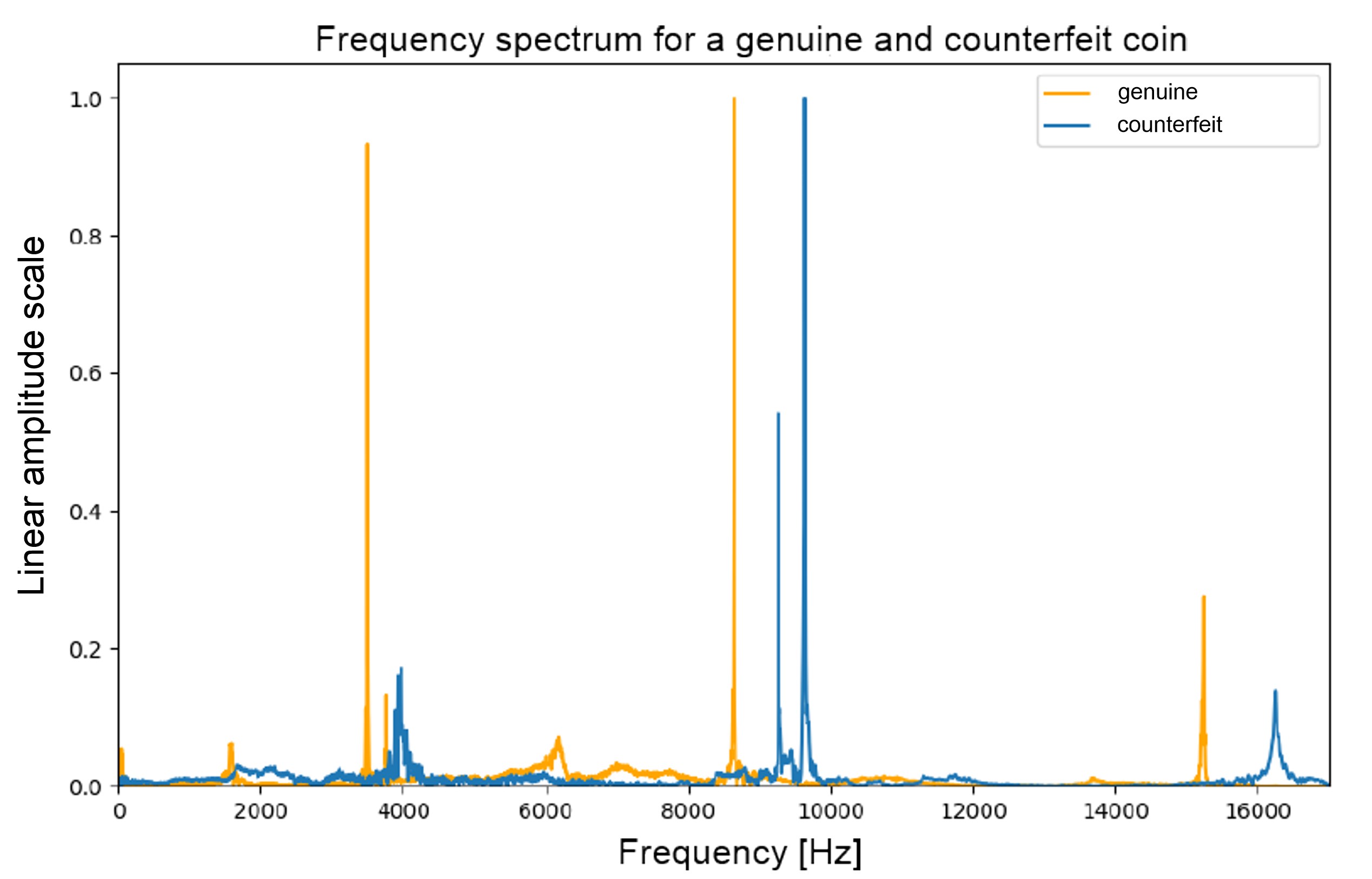}
    \caption{Frequency spectrum for genuine and counterfeit coins}
    \label{img_org_fake}
\end{figure}

Figure \ref{img_org_fake} compares the frequency spectra of an authentic coin and its counterfeit. The counterfeit is characterized by identical external dimensions and visual resemblance to the original. The author was unable to distinguish it from the original solely by visual inspection. The counterfeit coin, however, consists of a tungsten core covered with a thin silver layer. This combination of materials leads to significant changes in the coin's acoustic and mechanical properties, which, in turn, affect its resonance frequencies and natural vibration amplitudes. This is primarily due to differences in the density of silver and tungsten, Young's modulus, and Poisson's ratio. As a result, the frequency spectrum of the counterfeit coin shows a significant deviation from the authentic coin presented in the same graph. Shifts between peaks, changes in their amplitude, and even their absence are visible.

\subsection{Anomaly detection}

The development of machine learning models for the acoustic analysis of coins is fundamentally constrained by the limited availability of large, diverse training datasets. In contrast to domains such as image or speech recognition, publicly accessible repositories containing acoustic responses of coins are virtually nonexistent. Consequently, the dataset utilized in this study consists exclusively of recordings acquired by the authors, comprising only several to a dozen samples per coin type. Such a limited sample size is insufficient for effective training of deep learning models, which typically require thousands of instances to achieve robust generalization and to mitigate overfitting.

An additional challenge arises from the complete absence of data representing counterfeit coins. Due to legal, practical, and logistical constraints, acquiring verified counterfeit specimens is highly difficult, if not infeasible. As a result, the dataset is inherently incomplete, containing only authentic examples. This limitation precludes the direct application of supervised learning methods to binary classification tasks, such as distinguishing genuine from counterfeit coins. This scenario is characteristic of specialized research domains, where empirical data collection is both costly and time-consuming.

To address these limitations, several methodological strategies can be employed, particularly those tailored for learning from small or imbalanced datasets. A widely adopted approach is data augmentation, which enables artificial expansion of the dataset without the need for additional measurements \cite{SK19Augment}. In the context of acoustic signals, augmentation techniques include adding Gaussian noise, temporal shifting, and simulating varying acoustic environments. These transformations increase sample diversity and improve model robustness.

Another viable approach is transfer learning, which leverages models pre-trained on large-scale datasets for related tasks \cite{WKW16Transfer}. For instance, convolutional neural networks originally developed for image recognition can be adapted to process time-frequency representations of acoustic signals, such as spectrograms \cite{Boa16Signal}. By fine-tuning only the final layers, the requirement for extensive domain-specific data is significantly reduced.

Given the absence of counterfeit data, anomaly detection methods constitute a particularly relevant alternative. Techniques such as autoencoders \cite{hinton2006reducing, vincent2010stacked, sakurada2014anomaly} or one-class classification models \cite{scholkopf2001estimating} can be trained exclusively on authentic coin responses, learning a representation of normal acoustic behavior. During inference, deviations from this learned representation are interpreted as potential anomalies, corresponding to counterfeit specimens. However, this approach introduces additional challenges, including threshold selection and sensitivity to noise \cite{Recan2021Pred}, \cite{Vahdat2020NVAE}.

Autoencoders constitute a fundamental class of neural network models for unsupervised representation learning. Deep autoencoders can learn compact latent representations that preserve the essential structure of high-dimensional data \cite{hinton2006reducing}. This approach laid the groundwork for numerous extensions that improve robustness, regularization, and generative capabilities.

Autoencoders have also been widely applied to anomaly detection and one-class classification problems. In such settings, the model is trained exclusively on data from a target (normal) class and is expected to poorly reconstruct out-of-distribution samples \cite{sakurada2014anomaly}. The underlying assumption is that the learned latent manifold captures the distribution of normal data, while anomalous samples lie outside this manifold and thus yield higher reconstruction errors.

More broadly, one-class classification methods aim to distinguish normal data from all other possible inputs without explicit access to negative examples. Classical approaches, such as one-class support vector machines \cite{scholkopf2001estimating, ruff2018deep}, define decision boundaries in feature space, whereas deep learning-based methods leverage representation learning to model complex data distributions. Approaches based on the autoencoders are particularly attractive due to their simplicity, scalability, and ability to operate in a fully unsupervised manner. As a result, they have been successfully applied across various domains, including industrial inspection, fault detection, and acoustic signal analysis.

In summary, the literature demonstrates that autoencoders and their variants provide a flexible and theoretically grounded framework for representation learning and anomaly detection. Their ability to model the intrinsic structure of data without requiring labeled negative samples makes them especially suitable for one-class classification problems, including applications involving acoustic signatures of physical objects \cite{bulusu2021review}.

\subsection{Data generation}

Synthetic data generation offers a complementary solution. Generative models, including generative adversarial networks \cite{Pang2021Anomaly}, can be employed to produce artificial acoustic responses that approximate real measurements. Furthermore, physics-based simulations, for example, using finite element analysis tools such as COMSOL Multiphysics \cite{COMSOL2023Reference}, enable modeling the vibrational behavior of coins with varying material properties. By incorporating parameter differences, such as density or elasticity, it is possible to simulate the resonance characteristics of counterfeit coins, thereby partially compensating for the lack of empirical data.

Despite these mitigation strategies, the dataset remains inherently imbalanced, with a strong dominance of normal, i.e., authentic, samples and a scarcity or complete absence of anomalous cases. This imbalance leads to a bias in model training, often resulting in increased rates of false positives, where genuine coins are incorrectly classified as anomalies, or false negatives, where counterfeit coins are not detected \cite{SK19Augment}. In the context of acoustic coin analysis, anomaly detection is particularly challenging due to the lack of ground-truth data for counterfeit specimens. The resonance spectra of such coins are difficult to obtain experimentally, which further exacerbates the risk of misclassification.

\section{Autoencoder applications in counterfeit coin detection}
One particularly promising solution is the use of autoencoders, or neural networks based on an encoder-decoder architecture. These networks learn to reconstruct input data, minimizing reconstruction error based solely on authentic data \cite{TBL18Autoencoder}.
In the case of the presented system, an autoencoder trained on the frequency spectra of authentic coins can reconstruct counterfeit values in the test data. This simple approach, however, has the drawback of requiring the determination of a reconstruction error threshold. High values of this threshold will indicate anomalies in the input data, i.e., material or geometric defects. This approach effectively addresses the lack of data on counterfeit coins, enabling unsupervised classification based solely on authentic samples and making it an ideal tool for the presented solution.

\subsection{Autoencoder}

An autoencoder is an unsupervised neural network that learns an efficient data representation by minimizing reconstruction error. Their basic architecture consists of three main components

\begin{itemize}
  \item an encoder,
  \item an intermediate representation (latent space),
  \item a decoder.
\end{itemize}

The encoder maps input data $x \in R_d$ to a compressed representation $z \in R_m$, where $m < d$ is typically the case, allowing for dimensionality reduction and feature extraction [26]. This process is described by the encoding function $z = f_{\phi}(x)$, where $\phi$ represents the network parameters (weights and biases). The latent space $z$ stores the encoded information, which is then decoded into a reconstruction $\hat{x} = g_{\phi(z)}$, with parameters $\phi$. The entire model minimizes a loss function, typically the mean square error (MSE), expressed as (\ref{eq_w1})

\begin{equation}
 \mathcal{L}(\mathbf{x},\hat{\mathbf{x}}) = \frac{1}{n} \sum_{i=1}^{n} \| \mathbf{x} - \hat{\mathbf{x}} \|^2 
 \label{eq_w1}
 \end{equation}

This task involves reconstructing the input data in an autoassociative network using a compressed space containing the input's hidden features. This is a lossy compression method that omits irrelevant features, retaining only the most important ones, describing the most important characteristics of the process or model being studied.

An autoencoder can be effectively employed for single-class classification, where only data from the target class are available during training. In this paradigm, the autoencoder model is optimized to learn a compact latent representation that captures the intrinsic structure and distribution of the normal class by minimizing reconstruction error. Because the autoencoder is exposed exclusively to samples from this class, it becomes highly specialized in accurately reconstructing similar inputs, while exhibiting significantly higher reconstruction errors for samples that deviate from the learned manifold. This discrepancy provides a natural decision criterion, where a threshold on the reconstruction error can be used to distinguish in-class instances from outliers. 

The autoencoder training process uses backpropagation with gradient optimization, such as the Adam or SGD algorithms. During iterations, the network adjusts its parameters to minimize the reconstruction loss, thereby learning a compressed representation. Internally, the encoder and decoder layers consist of neurons with nonlinear activation functions.

For acoustic data, such as coin frequency spectra, the encoder can extract features, including peak frequencies and amplitudes, while the decoder reconstructs the original signal, enabling anomaly detection by applying a reconstruction error threshold. Large values of this error indicate deviations from the original class, such as counterfeit coins made from different materials or with different geometric properties.

\section{Implementation of the developed solution}

Two models were developed: an autoencoder for authenticity verification and a classifier for coin type recognition. These models were designed for analyzing audio recordings, so appropriate preparation of the input data was required.

\subsection{Acoustic data preparation}

Recording normalization and processing were required to enable training on recordings obtained from various sources. The recordings were standardized, and augmentation was used to artificially increase the amount of data, thus increasing the robustness of the trained models to varying recording conditions.

A potential recording created by the user could be long and contain multiple impacts. A segment of the recording immediately after the coin impact is extracted to minimize the sound's impact on the trained neural network models. Additionally, initial silence and the excess signal after the coin resonance have ceased. This process begins by calculating a detection threshold, empirically determined to be 10\% of the maximum amplitude.

The maximum amplitude is calculated from the signal's absolute value. Finally, the signal's onset is set to the point at which the amplitude exceeds the previously calculated threshold. To minimize the impact on the model's results, the segment's starting position is shifted by a constant time value (default: 0.08 s).

\begin{figure}
    \centering
    \includegraphics[width=1\linewidth]{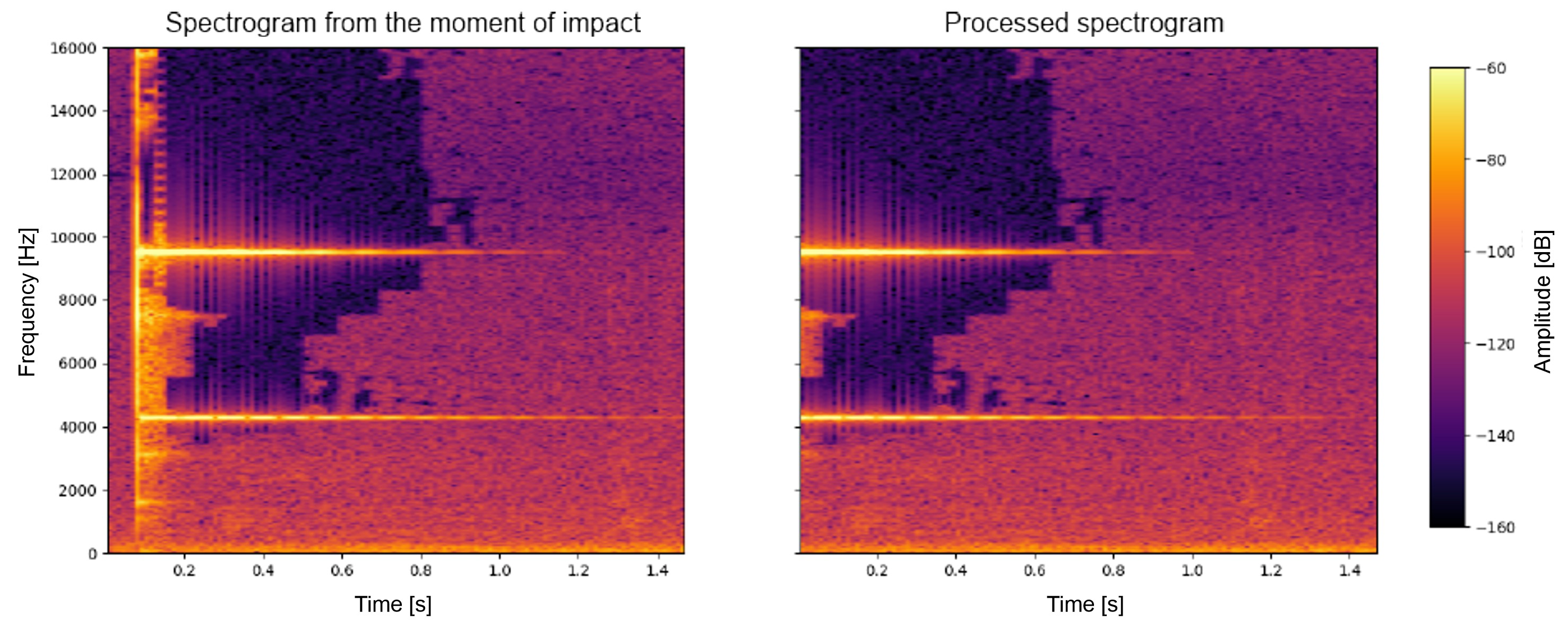}
    \caption{Example of a sound spectrogram before impact (left) and after it has been modified by the algorithm (right)}
    \label{img_spectr_proc}
\end{figure}

If the signal length is insufficient, zero padding is applied to ensure a constant length for the extracted segment and avoid unnecessary noise. The final recording length was 8820. A sample recording of a coin impact, extracted by the implemented algorithm for further processing, is shown in Figure~\ref{img_spectr_proc}.

\subsection{Normalization of loudness levels}

Root mean square (RMS) normalization is used to standardize the loudness level of recorded audio signals from struck coins. Because recordings can be made at different microphone distances from the sound source, or because the equipment's sensitivity may vary, the resulting amplitude differences can negatively impact the training efficiency of neural network models. Normalization ensures consistency of input data, increasing the models' robustness to differences in audio recording conditions. It also improves the models' generalization ability.

The normalization process involves calculating the RMS value for the clipped signal (\ref{eq_w2})

\begin{equation}
\mathrm{RMS}=\sqrt{\frac{1}{N}\sum_{n=1}^{N}s[n]^2}
\label{eq_w2}
\end{equation}

where
\begin{itemize}
\item $N$ - number of samples in the signal segment,
\item $s[n]$ - signal value in the $n$-th sample.
\end{itemize}

The signal is then scaled to achieve the target RMS value (default $0.1$), which is given by the formula (\ref{eq_w3})

\begin{equation}
 s_{\mathrm{norm}}[n]=\frac{s[n]}{\mathrm{RMS}}\cdot \mathrm{RMS}_{\mathrm{target}}
\label{eq_w3}
\end{equation}

This ensures that all processed signals have a similar energy level, facilitating further spectral analysis and model training.

\subsection{Data Augmentation}

Data augmentation increases the diversity of a training dataset by modifying existing data. It improves the robustness of neural network models, where the underlying dataset may not account for the variability of user-generated recording conditions.

In audio signals, variability can arise from differences in signal amplitude and from background noise or other interference.

In the case analyzed, augmentation involves two steps:

\begin{itemize} 
\item changing the signal amplitude after RMS normalization,
\item adding Gaussian noise.
\end{itemize}

In the first step, the signal is scaled by multiplying it by various amplitude factors. This allows for the generation of signal variants with different loudness levels while maintaining the normalized RMS energy.

In the next step, Gaussian noise with zero expectation and a specified standard deviation is added to the signal. This noise simulates natural interference that may occur in real recordings, such as microphone noise or background noise. Mathematically, this process can be described as (\ref{eq_w4})

\begin{equation}
s_{\mathrm{aug}}[n] = \mathrm{coeff} \cdot 
s_{\mathrm{norm}}[n] + \eta[n]
\label{eq_w4}
\end{equation}

where
\begin{itemize}
\item $s_{\mathrm{aug}}$ - augmented signal in the $n$-th sample,
\item $\mathrm{coeff}$ - amplitude scaling factor,
\item $s_{\mathrm{norm}}[n]$ - normalized signal after RMS normalization,
\item $\eta[n]$ - Gaussian noise with a distribution of $N(0, \sigma)$, where $\sigma$ is the noise standard deviation.
\end{itemize}

After augmentation, the signal is further processed with a Hamming window to minimize spectral leakage during frequency analysis [38]. The frequency spectrum is then computed using a fast Fourier transform (FFT), and the spectrum is normalized by the maximum amplitude to ensure consistency between the input data and the trained model.

\subsection{Spectrum Peak Search}

To search for peaks in the spectra, an algorithm was used to detect values that are elevated above their neighbors, based on a dynamic prominence threshold (a measure of how much a given peak "clearly" stands out from its neighbors). Prominence is calculated as five times the median absolute deviation. This allows peak search to adapt to the noise level in the spectrum [40]. The minimum peak height is 15\% of the maximum amplitude,
while the minimum distance between peaks is 150 frequency samples.

Once all peaks are found, the 10 largest in amplitude are selected, allowing for focusing on the dominant harmonics. Once all peaks are found, the distance between the sets of peaks in the original and reconstructed spectra is calculated. To calculate the distance, the closest peak from the reconstructed spectrum is assigned to each peak from the original spectrum. Ultimately, this distance is calculated using the formula \ref{eq_w6}.

\begin{equation}
d=\sqrt{w_f(f_1-f_2)^2+w_a(a_1-a_2)^2}
\label{eq_w6}
\end{equation}

where:
\begin{itemize}
    \item $f_1$, $f_2$ - frequencies of the original and reconstructed peaks,
   \item $a_1$, $a_2$ - amplitudes of the original and reconstructed peaks,
   \item $w_f = 2.0$, $w_a = 0.5$ - weights for frequency and amplitude, respectively.
\end{itemize}

The total distance is the sum of the minimum distances for matched peak pairs, and an additional penalty mechanism is added if the sets contain mismatched peaks.

The algorithm's threshold was experimentally set to 131.

\section{Neural model}

In the implemented application, the autoencoder is a neural network used to verify a coin's authenticity by analyzing its frequency spectrum. Using a model trained only on recordings of genuine coins means it is trained to reconstruct only real coins. This ensures that when reconstructing a recording of a real coin, the reconstruction is accurate, resulting in a very small reconstruction error. If the recording is from a different coin, unknown to the trained model, or if the coin is counterfeit, the reconstruction error will be much larger. Finally, by algorithmically setting an error threshold, the application can assess the authenticity of the verified coin.

The implemented autoencoder consists of two parts: an encoder and a decoder. The encoder accepts input data (in the case of the application being developed, the frequency spectrum) and, in the next step, compresses it to a lower dimensionality, known as an intermediate representation. It then passes it to the decoder. At this point, the decoder attempts to reconstruct the original signal spectrum from the provided representation, and after this step, the reconstructed output signal is compared to the input signal.

In the implemented model, the encoder consists of three linear layers with ReLU activation functions and random neuron dropout for regularization, i.e., limiting the model's complexity to prevent overfitting. The subsequent layers of the encoder are presented below:

\begin{itemize}
    \item Input layer (feature vector) with a length of 8820 values.
    \item Hidden layer of 1024 neurons with ReLU activation and random neuron dropout set to 0.1.
    \item Hidden layer of 512 neurons with ReLU activation.
    \item Representation layer of 128 neurons. The output signals of this layer are the characteristics of the coins' sound responses. The autoencoder used is designed to detect differences between the characteristics of genuine coins and counterfeits.
\end{itemize}

The decoder is symmetric with respect to the encoder and consists of the following layers:

\begin{itemize}
    \item A hidden layer of 512 neurons with ReLU activation,
    \item A hidden layer of 1024 neurons with ReLU activation and random neuron exclusion set to 0.1.
    \item An output layer of 8820 values.
\end{itemize}

The autoencoder training process was based on minimizing reconstruction error, measured by mean squared error (MSE). The Adam optimizer is used with a learning rate of 0.0005, and training is performed for 30 epochs with a batch size of 4. After training, an anomaly detection threshold is calculated.

\subsection{Anomaly Detection Threshold}

The detection threshold is a numerical value representing the maximum deviation in the frequency spectrum of a coin reconstructed by the autoencoder before the coin is classified as counterfeit. This threshold is determined dynamically using a trained autoencoder and the training data used to train it. The algorithm determines the threshold statistically based on the distribution of distances between matched peaks in the original and reconstructed spectra. The threshold formula is presented in (\ref{eq_w5})

\begin{equation}
\mathrm{threshold} = \mu_d + 3\sigma_d
\label{eq_w5}
\end{equation}

where
\begin{itemize}
\item
\item $\mu_d$ - the mean distance between matched peaks,
\item $\sigma_d$ - the standard deviation of the distance between matched peaks of sound.
\end{itemize}

Adding three times the standard deviation ensures that 99.75\% of observations are within this range.

\subsection{Classifier}

To recognize coin types based on their characteristic resonance frequencies, a classifier was developed using a latent feature space (intermediate autoencoder representation). This representation is created from the frequency spectrum by an encoder operating within the autoencoder model. 

Using an intermediate representation for classification is an effective approach because it extracts key spectral features while omitting noise and less important details. This leads to high precision in distinguishing between the given classes without the need for manual feature extraction. Additionally, the autoencoder generates a representation that generalizes various recording conditions. This representation is augmented to account for differences in microphone distance and background noise levels.

\begin{figure}[ht]
    \centering
    \includegraphics[width=1\linewidth]{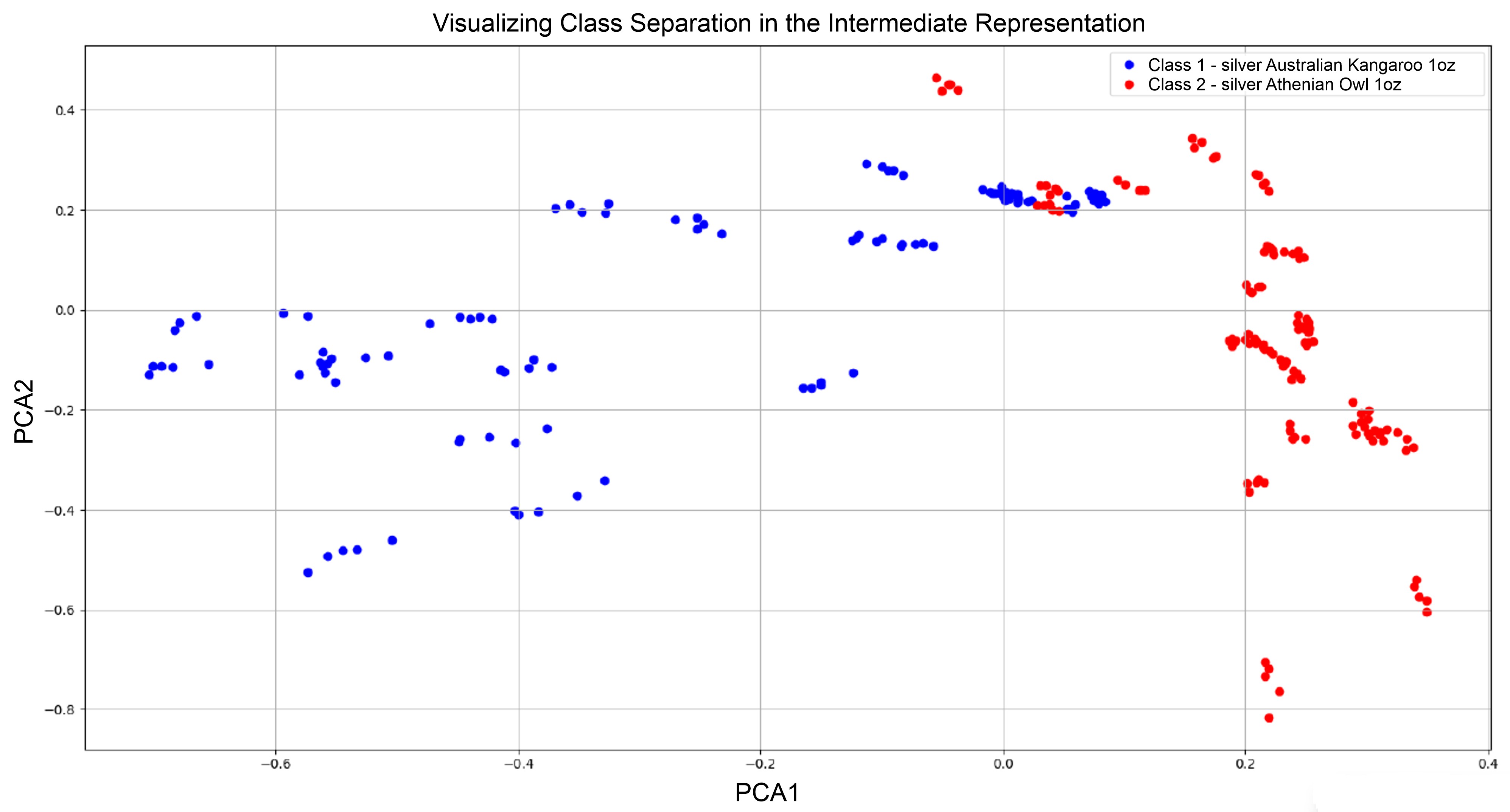}
    \caption{PCA visualization of the autoencoder intermediate representation for two types of coins}
    \label{img_pca}
\end{figure}

Figure \ref{img_pca} presents a visualization of the sample distribution in the intermediate representation, obtained from the output of an encoder trained on two types of silver coins (Australian Kangaroo and Athenian Owl \cite{MK21Awers, MP_Ovl}). The principal component analysis (PCA) technique was used to reduce the dimensionality of the encoder output (from 128) to 2 in order to plot the sound response distribution \cite{GGH22PCA}.

The diagram shows a separation, especially along the first principal component, indicating that the autoencoder's feature space allows differentiation among individual classes. Information was likely lost during dimensionality reduction, which is why the diagram shows a partial mix of classes. Despite this, this is not a problem, as the classifier performs well in class recognition.

Due to the good linear separation of extracted features in the autoencoder's hidden layer, a two-layer classifier proved sufficient: 64 neurons with ReLU functions in the first layer and two linear neurons in the output layer. The training data consisted of features from two classes. The classifier's task was to assign the input data to one of the two classes (Australian Kangaroo or Athenian Owl) or to mark it as an unrecognized coin (indicating a counterfeit).

Cross entropy (CE) was used as the loss function. The Adam optimizer was used with a learning rate of 0.0005, 30 epochs, and a batch size of 4. During training, the loss and classification accuracy were measured.

\section{Experimental results}

\begin{figure}[ht]
    \centering
    \includegraphics[width=1\linewidth]{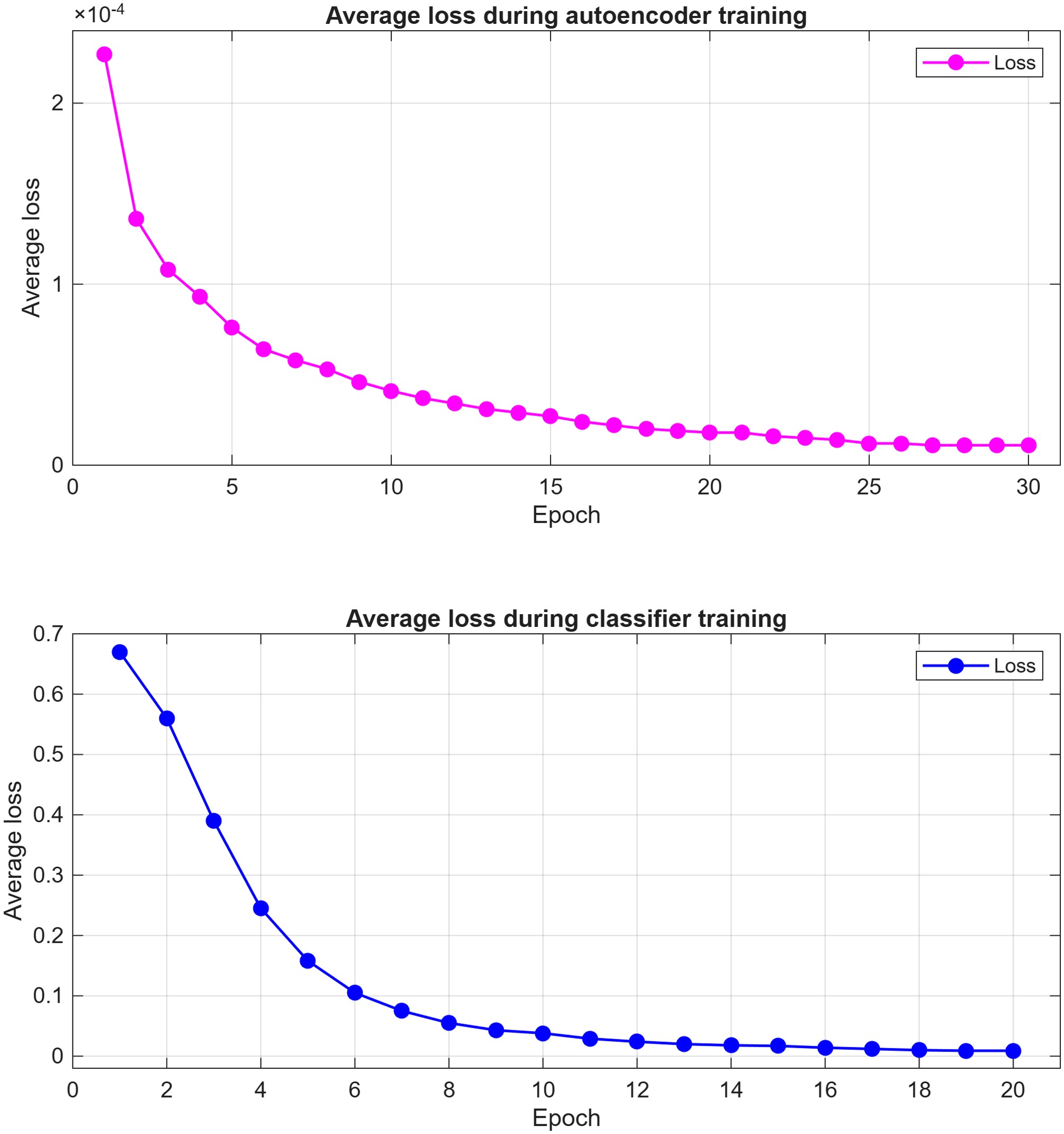}
    \caption{Autoencoder and classifier training curves}
    \label{img_learn}
\end{figure}

Figure \ref{img_learn} shows the average loss over successive epochs during training of the autoencoder model. The loss function is the mean squared error (MSE) between the reconstructed and original spectra. 

The graph shows a decrease in loss over successive epochs. It is worth noting that the mean square error for the entire spectrum is immediately small because this measure considers the entire frequency spectrum, where most values are close to zero. This is because the coin's sound spectrum is dominated by high-amplitude resonance peaks, while the rest of the spectrum is either noise or near zero. Therefore, the MSE error is low. Although the model was trained on a small amount of data (only two classes, lacking diversity), it stabilized its loss after approximately 25 epochs while maintaining adequate generalization.

\section{Coin Verification Results}

\begin{figure}[ht]
    \centering
    \includegraphics[width=1\linewidth]{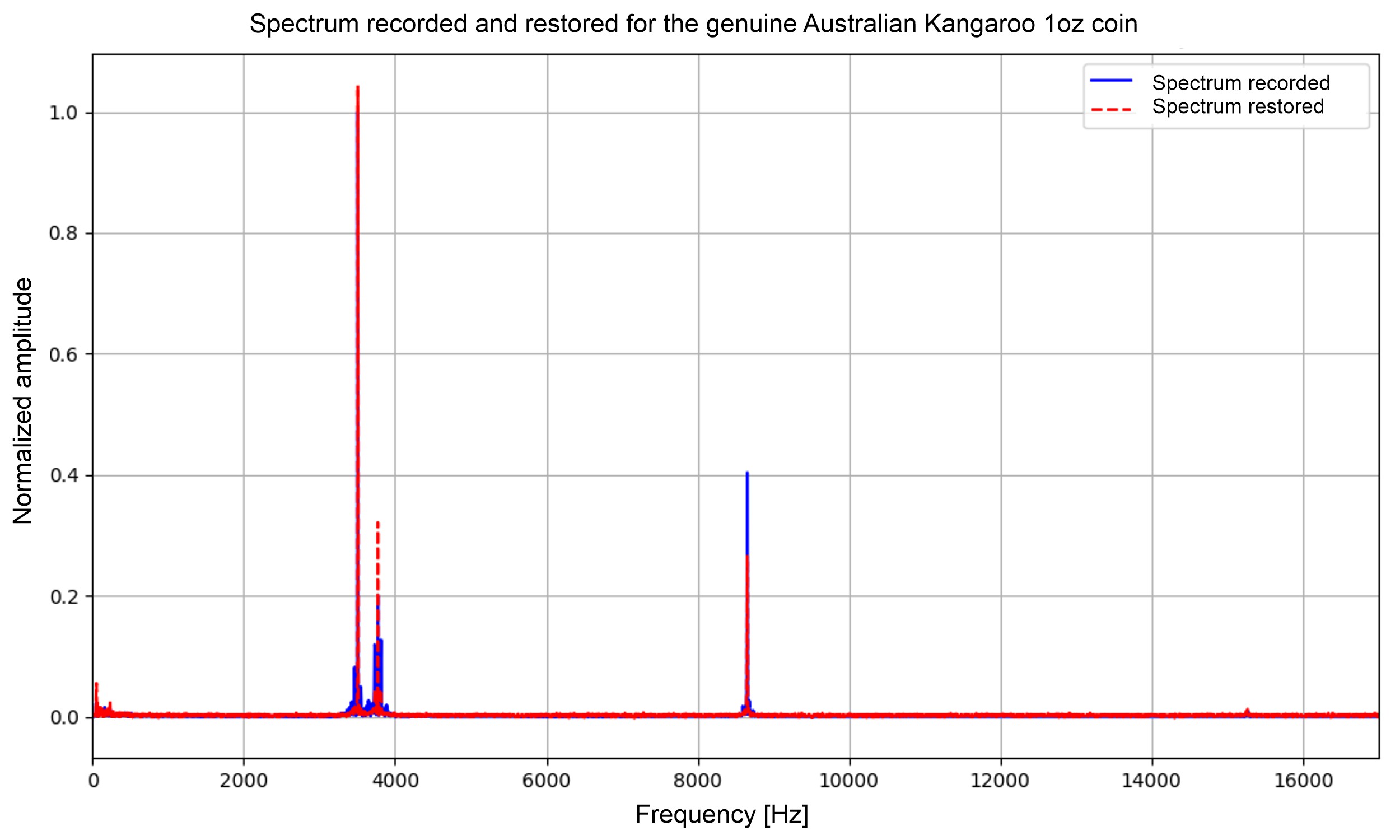}
    \caption{Spectrum recorded and restored for the genuine Australian Kangaroo 1 oz coin}
    \label{img_peak_genui}
\end{figure}

Figure \ref{img_peak_genui} shows a comparison of the original recording of the impact of a 1 oz Australian Kangaroo silver coin with the spectrum reconstructed by the autoencoder. The horizontal axis is the frequency in the spectrum, limited to 0-16 kHz, while the vertical axis is the normalized amplitude, ranging from 0 to 1. 

A high degree of similarity is evident between the two spectra. The original spectrum shows the largest peak at around 3510 Hz, which the autoencoder perfectly reconstructed. The next two peaks, around 3760 Hz and 8650 Hz, were also reconstructed, but their normalized amplitude differs from the original. However, this is not a problem, as the calculated distance between the peaks takes frequency into account more, it accounts for 80\% of the result, whereas amplitude only accounts for 20\% of the calculated distance. This yielded a final peak distance of 0.13, which exceeds the threshold for recognizing a counterfeit coin (131), indicating a positive authentication of the coin. Additionally, the classifier confirmed the identified coin type as a 1 oz silver Australian Kangaroo.

The second test was performed on a counterfeit 1 oz silver Australian Kangaroo coin made with a tungsten core. The coin is visually very similar to its original counterpart and is extremely difficult to distinguish at a glance.

\begin{figure}[ht]
    \centering
    \includegraphics[width=1\linewidth]{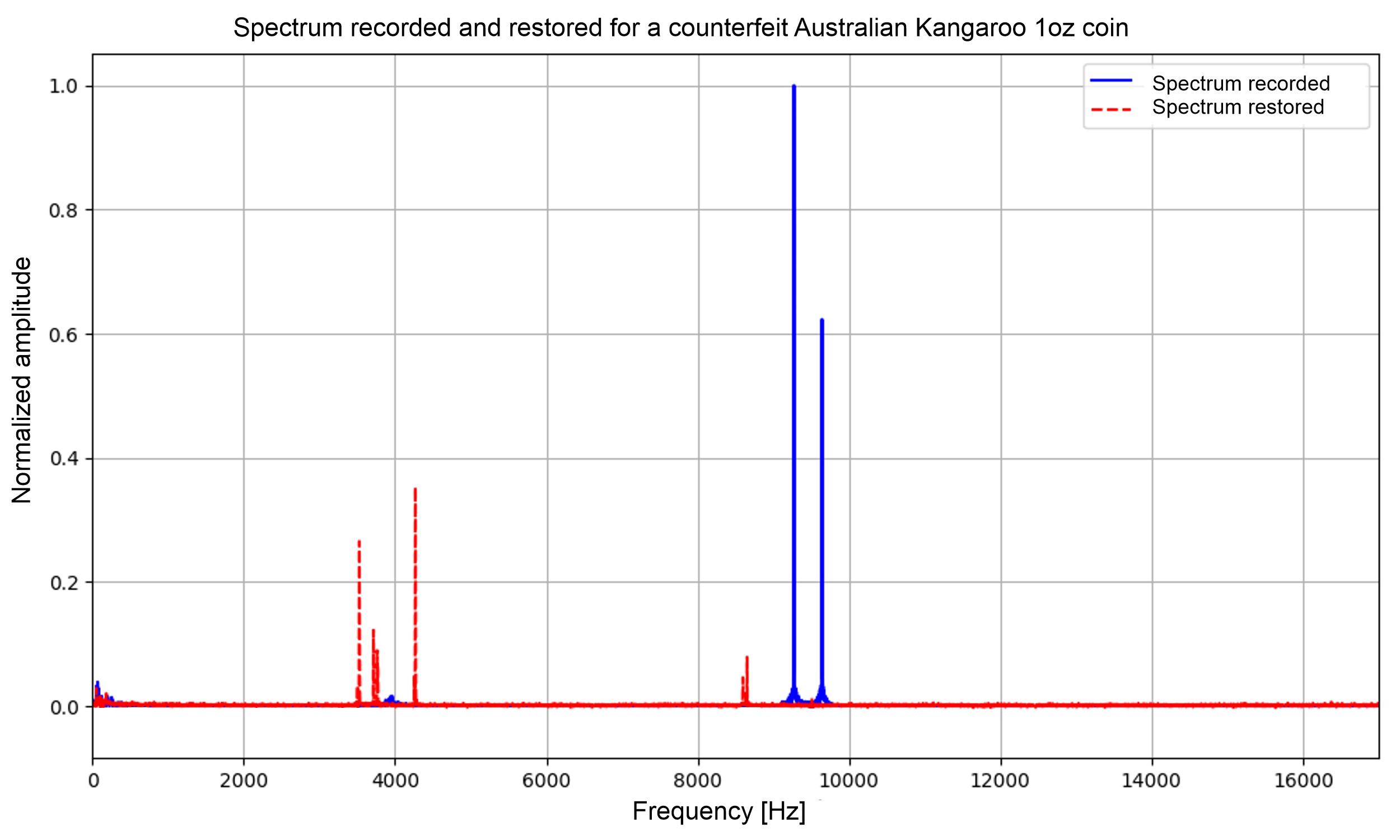}
    \caption{Spectrum recorded and restored for a counterfeit Australian Kangaroo 1 oz coin}
    \label{img_peak_fake}
\end{figure}

Figure \ref{img_peak_fake} shows a drastic difference in the input spectrum compared to the original coin presented in \ref{img_peak_genui}. It is obvious that this change in the coin's characteristics is caused by the changed material properties of the coin being tested.

Because the autoencoder wasn't trained to reproduce the characteristics of a counterfeit coin, it can't reconstruct its input spectrum from its intermediate representation. This results in a large discrepancy in the reconstructed spectrum, with the reconstructed peaks not overlapping with the input peaks at all. This causes the algorithm to assign penalties during the distance calculation, resulting in a final distance of approximately 1077. This distance is significantly above the assumed 131 threshold, so this coin is marked as counterfeit (unrecognized) and therefore not subject to further classification.

The next study examined the original silver Athenian Owl 1 oz coin, which was part of the autoencoder's training set. The reconstructed frequency spectrum matches the input exactly. The sum of the peak spacing is 0.04, so the coin is considered authentic, and the classifier then recognizes its type.

\begin{figure}
    \centering
    \includegraphics[width=1\linewidth]{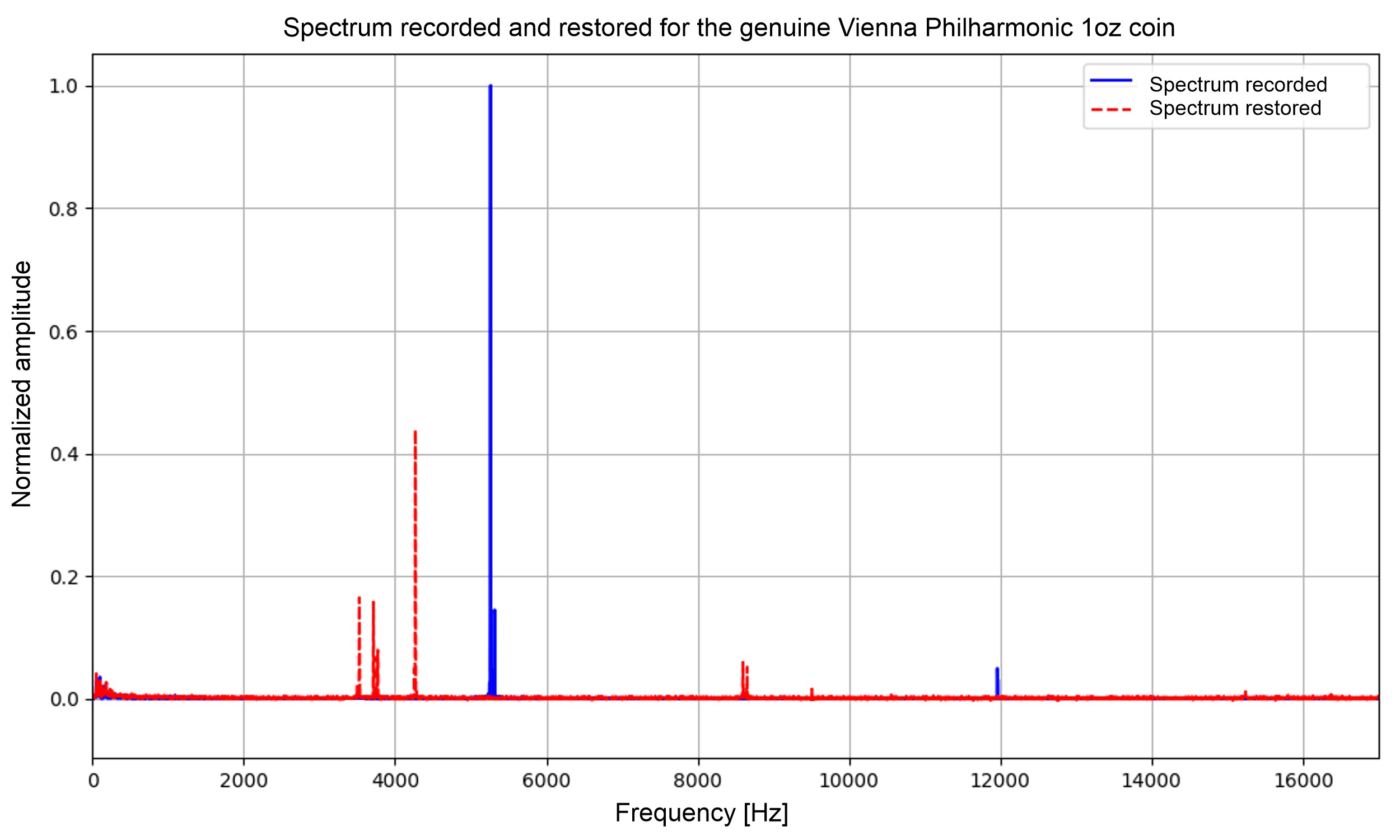}
    \caption{Spectrum recorded and restored for the genuine Vienna Philharmonic 1 oz coin}
    \label{img_peak_genui_vienna}
\end{figure}

The next test involved examining a genuine coin, a Vienna Philharmonic 1 oz \cite{MP_filh}, that was not part of the autoencoder training data. Although it is a real coin, the neural network is unaware of it (it was not included in the training dataset), leading to an incorrect reconstruction of the spectrum. The model incorrectly reproduces the encoded intermediate representation, resulting in a final total peak distance of approximately 1708. The coin is classified as counterfeit (unrecognized by the model), even though it is genuine (Figure \ref{img_peak_genui_vienna}).

\section{Conclusions}

A primary strength of the proposed solution lies in the synergy between the autoencoder and the classifier models. By operating in tandem, these architectures provide a comprehensive authentication framework. The autoencoder, trained exclusively on authentic specimens, serves as an anomaly detector to differentiate genuine coins from counterfeits. Simultaneously, the classifier provides granular data regarding the specific coin type. Both models demonstrated rapid convergence during training and strong generalization on unseen test samples.

This integrated approach enables non-destructive verification across various coin types. A key technical innovation is the implementation of a dynamically calculated anomaly threshold. This threshold enables robust rejection of counterfeits or unknown specimens, especially when reconstruction errors exceed the baseline by an order of magnitude. Consequently, the combination of spectral reconstruction and dynamic thresholding creates a stable verification environment that remains resilient against variations in recording devices or environmental conditions.

While the current models achieve high precision, they are sensitive to novel data, a direct consequence of the non-diverse, relatively constrained dataset used for training. Currently, the system adopts a conservative security posture: any coin type unknown to the autoencoder is categorized as an anomaly (counterfeit).

The challenges regarding the dataset are twofold. First, high-quality recordings of diverse bullion coins are currently difficult to access, as manually compiling such a database is both cost-prohibitive and time-intensive. Second, the models were trained on a limited number of classes, necessitating further expansion to include a broader array of minting series. 

To mitigate these limitations, data augmentation was employed to artificially enhance the diversity of the training set. This was particularly effective because the original recordings were conducted in both controlled (quiet) and high-noise environments. This dual-condition recording strategy ensures that the augmentation process reflects actual testing performance and environmental interference. However, while augmentation provides a necessary stopgap, the long-term stability and scalability of the solution will require integrating a larger volume of authentic, real-world training data.

\subsection{Future Work and Research Directions}

Future research will focus on advancing the current architecture to overcome data limitations and enhance classification accuracy. A primary objective is the implementation of a deep variational autoencoder (VAE), which offers a more sophisticated probabilistic approach to modeling the unique acoustic signatures that serve as a coin's physical fingerprint. Additionally, we plan to evaluate various modern deep learning architectures to establish a more robust benchmark for non-destructive verification. 

To specifically mitigate the challenges associated with small datasets, future iterations will explore the application of "tiny" Large Language Models (LLMs). By utilizing these models in zero-shot and few-shot prompting modes, we aim to achieve high-precision authentication even with minimal training instances. This approach significantly reduces reliance on the expensive resource data collection.

\bibliographystyle{plain} 
\bibliography{arxiv_bulion}

\end{document}